\pdfoutput=1 
\documentclass[conference]{IEEEtran}

\usepackage{amsmath,amssymb,amsfonts}
\usepackage{algorithmic}
\usepackage{graphicx}
\def\BibTeX{{\rm B\kern-.05em{\sc i\kern-.025em b}\kern-.08em
    T\kern-.1667em\lower.7ex\hbox{E}\kern-.125emX}}

\usepackage{tabularx}
\usepackage{listings}
\usepackage{multirow}
\usepackage{numprint}
\usepackage[table,dvipsnames]{xcolor}
\usepackage{xspace}
\usepackage{floatrow}
\usepackage[caption=false]{subfig}
\usepackage[style=base]{caption}
\usepackage{transparent}
\usepackage[normalem]{ulem}
\ifdefined\RELEASE
\usepackage[sort,nocompress]{cite}
\else
\usepackage{fancyhdr}
\usepackage[sort,nocompress]{cite}
\usepackage[bookmarks=true,breaklinks=true,colorlinks,linkcolor=blue,citecolor=blue,urlcolor=blue]{hyperref}
\fi
\definecolor{amber}{rgb}{1.0, 0.49, 0.0}
\definecolor{amethyst}{rgb}{0.6, 0.4, 0.8}
\newcommand{\squishlistnum}{  %
 \newcounter{qcounter}
 \begin{list}{\roman{qcounter})~}{\usecounter{qcounter}}
  { \setlength{\itemsep}{0pt}
     \setlength{\parsep}{0pt}
     \setlength{\topsep}{0pt}
     \setlength{\partopsep}{0pt}
     \setlength{\leftmargin}{0em}
     \setlength{\labelwidth}{0em}
     \setlength{\labelsep}{0em} } }

\newcommand{\squishlist}{ %
 \begin{list}{$\bullet$}
  { \setlength{\itemsep}{2pt}
     \setlength{\parsep}{0pt}
     \setlength{\topsep}{2pt}
     \setlength{\partopsep}{0pt}
     \setlength{\leftmargin}{1em}
     \setlength{\labelwidth}{1em}
     \setlength{\labelsep}{0.5em} } }

\newcommand{\squishend}{
  \end{list}  }
  
\newcommand{\spotnospace}{DBG\xspace}
\newcommand{\spot}{\spotnospace}
\newcommand{\spotacronym}{{\em \underline{D}egree-\underline{B}ased \underline{G}rouping (\spotnospace)}\xspace}
\newcommand{\spotfullname}{Degree-Based Grouping\xspace}

\newcommand{\dbg}{\spot}
\newcommand{\dbgacronym}{\spotacronym}
\newcommand{\dbgfullname}{\spotfullname}
\newcommand\stitchref[2]{\ref{#1}\subref{#2}}

\newcommand\noindentsection[1]{\vspace{0.2\baselineskip}\noindent}
\newcommand\noindentsectiontitle[1]{\vspace{0.2\baselineskip}\noindent {\em #1}}

\newcommand\visiblespace{\vspace{0.2\baselineskip}}

\newcommand\visiblespacel{\vspace{0.3\baselineskip}}
\newcommand\visiblespacexl{\vspace{0.4\baselineskip}}

\newcolumntype{a}{>{\columncolor{Gray}}r}
\newcolumntype{b}{>{\columncolor{white}}r}
\definecolor{LightCyan}{rgb}{0.88,1,1}

\newcommand\tw{\emph{tw}\xspace}
\newcommand\fr{\emph{fr}\xspace}
\newcommand\wl{\emph{wl}\xspace}
\newcommand\kr{\emph{kr}\xspace}
\newcommand\lj{\emph{lj}\xspace}
\newcommand\pl{\emph{pl}\xspace}

\newcommand\sd{\emph{sd}\xspace}
\newcommand\mpi{\emph{mp}\xspace}

\newcommand\pr{PR\xspace}
\newcommand\prd{PRD\xspace}
\newcommand\sssp{SSSP\xspace}
\newcommand\bc{BC\xspace}

\newcommand\gorder{Gorder\xspace}

\newcommand\ie{{i.e., }}
\newcommand\eg{{e.g., }}

\newcommand\maxdegree{{$\mathbb{M}$}\xspace}
\newcommand\avgdegree{{$\mathbb{A}$}\xspace} %
\newcommand\constant{{$\mathbb{C}$}\xspace} %
\newcommand\infinity{{$\infty$}\xspace}
\usepackage{pifont}%
\newcommand{\cmark}{\ding{51}}%
\newcommand{\xmark}{\ding{55}}%
\newcommand\num[1]{\footnotesize{\numprint{#1}}}
\npthousandsep{,}\npthousandthpartsep{}\npdecimalsign{.}
\newcommand{\floor}[1]{\left\lfloor #1 \right\rfloor}

\usepackage{hyperref}
\hypersetup{
    colorlinks=true,
    filecolor=magenta,    
    bookmarks=true,
    breaklinks=true,
    colorlinks=true,
    linkcolor=blue,
    citecolor=ForestGreen,
    urlcolor=cyan,
    pdfauthor={Priyank Faldu et al.},
    pdftitle={A Closer Look at Lightweight Graph Reordering}%
}

\begin{document}

\title{A Closer Look at Lightweight Graph Reordering}%
\author{
\IEEEauthorblockN{Priyank Faldu}
\IEEEauthorblockA{The University of Edinburgh \\ \textit{priyank.faldu@ed.ac.uk}}
\and
\IEEEauthorblockN{Jeff Diamond}
\IEEEauthorblockA{Oracle Labs \\ \textit{jeff.diamond@oracle.com}}
\and
\IEEEauthorblockN{Boris Grot}
\IEEEauthorblockA{The University of Edinburgh\\ \textit{boris.grot@ed.ac.uk}}
}

\ifdefined\RELEASE
\else
\fancypagestyle{firstpage}{
    \fancyhf{}
    \renewcommand{\headrulewidth}{0pt}
    \fancyhead[C]{{\textbf{In Proceedings of the International Symposium on Workload Characterization (IISWC'19)}}}
    \pagenumbering{arabic}
}
\fi

\maketitle

\ifdefined\RELEASE
\else
\thispagestyle{firstpage}
\pagestyle{plain}
\fi

\begin{abstract}
Graph analytics power a range of applications in areas as diverse as finance, networking and business logistics. A common property of graphs used in the domain of graph analytics is a \emph{power-law distribution} of vertex connectivity, wherein a small number of vertices are responsible for a high fraction of all connections in the graph. These richly-connected \emph{(hot)} vertices inherently exhibit high reuse. However, their sparse distribution in memory leads to a severe underutilization of on-chip cache capacity. Prior works have proposed lightweight skew-aware vertex reordering that places hot vertices adjacent to each other in memory, reducing the cache footprint of hot vertices and thus improving cache efficiency. However, in doing so, they may inadvertently destroy the inherent community structure within the graph, which may negate the performance gains achieved from the reduced footprint of hot vertices. 

In this work, we study existing reordering techniques and demonstrate the inherent tension between reducing the cache footprint of hot vertices and preserving original graph structure. We quantify the potential performance loss due to disruption in graph structure for different graph datasets. We further 
show that reordering techniques that employ fine-grain reordering significantly increase misses in the higher level caches, even when they reduce misses in the last level cache.

To overcome the limitations of existing reordering techniques, we propose \emph{Degree-Based Grouping (DBG)}, a novel lightweight reordering technique that employs a coarse-grain reordering to largely preserve graph structure while reducing the cache footprint of hot vertices. Our evaluation on 40 combinations of various graph applications and datasets shows that, compared to a baseline with no reordering, DBG yields an average application speed-up of 16.8\% vs 11.6\% for the best-performing existing lightweight technique.
\end{abstract}


\section{Introduction}
\label{sec:intro}

Graph analytics is an exciting and rapidly growing field with applications spanning diverse areas such as optimizing routes, uncovering latent relationships, pinpointing influencers in social graphs and many more. Graph analytics exhibit highly irregular memory access patterns. As a result, when processing large graphs, graph analytics tend to lack both spatial and temporal locality, leading to frequent misses in on-chip caches, which limits their performance~\cite{locality-exists,imp,prefetch-data-structure,gorder,fc,hubcluster}.

A distinguishing property of graph datasets common in many graph-analytic applications is that the vertex degrees follow a skewed {\em power-law} distribution, in which a small fraction of vertices have many connections while the majority of vertices have relatively few connections~\cite{power-law,power-law-internet,powergraph,fc,hubcluster}.
Graphs characterized by such a distribution are known as {\em natural} or {\em scale-free} graphs and are prevalent in a variety of domains, including social networks, computer networks, financial networks, semantic networks and airline networks.

Power-law skew in the degree distribution means that a small set of vertices with the largest number of connections is responsible for a major share of memory accesses. The fact that these richly-connected vertices (henceforth referred to as {\em hot vertices}) comprise a small fraction of the overall footprint while exhibiting high reuse makes them prime candidates for caching. Meanwhile, the rest of the vertices (henceforth referred to as {\em cold vertices}) comprise a large fraction of the overall footprint while exhibiting low or no reuse.

For a typical graph application, each cache block (or line)
contains multiple vertices, as vertex properties usually require just 4 to 16 bytes whereas cache block size in modern processors is typically 64 or 128 bytes.
Since hot vertices are sparsely distributed throughout the memory space,
they inevitably share 
cache blocks with cold vertices, leading to underutilization of a considerable fraction of useful cache capacity.

Prior works have proposed lightweight reordering \mbox{techniques} that leverage application-visible graph properties, such as skew in vertex degree distribution, to improve cache block utilization~\cite{fc,hubcluster}.
We collectively refer to these techniques as {\em skew-aware} techniques. 
Skew-aware techniques reorder vertices in memory such that hot vertices are adjacent to each other in a contiguous memory region.
As a result, each cache block is comprised of exclusively hot or cold vertices, reducing the total footprint (\ie number of cache blocks) required to store hot vertices.
Blocks that are exclusively comprised of hot vertices are far more likely to be retained in the cache due to higher aggregate hit rates, leading to
higher utilization of existing cache capacity.

A straightforward way to pack vertices with similar degree into each cache block
is to
apply {\em Sort Reordering}, which sorts vertices in memory based on their degree.
However, Sort is not always beneficial, %
because many real-world graph datasets exhibit a strong internal structure, \eg clusters of webpages within the same domain in a web graph, or communities of common friends in a social graph~\cite{community1,community2}.
In such datasets, vertices within the same community that are accessed together often reside nearby in memory, exhibiting spatio-temporal locality that should be preserved. Fine-grain vertex reordering, such as Sort and Hub Sorting~\cite{fc}, destroys the spatio-temporal locality, which limits the effectiveness of such reordering techniques on graph datasets that exhibit structure. 

In this work, we quantify potential performance loss due to disruption of graph structure on various datasets.
We further characterize locality at all three levels of the cache hierarchy, and show that all skew-aware techniques are generally effective at reducing LLC misses. However, the techniques that employ fine-grain reordering significantly disrupt graph structure, which increases misses in higher level caches. %
Our results highlight a tension between reducing the cache footprint of hot vertices and preserving graph structure, which limits the effectiveness of existing skew-aware reordering techniques.

To overcome the limitations of existing techniques, we propose \dbgacronym, a novel reordering technique that largely preserves graph structure while 
reducing the cache footprint of hot vertices.
Like prior skew-aware techniques, \dbg segregates hot vertices from the cold ones. However, to preserve existing graph structure, \dbg employs {\em coarse-grain} reordering.
\dbg partitions vertices into a small number of groups based on their degree but maintains the original relative order of vertices within each group.
As \dbg \emph{does not sort} vertices within any group to minimize structure disruption, \dbg also incurs a very low reordering overhead.%

\visiblespacexl

To summarize, we make the following contributions:

\squishlist
    \item We study existing skew-aware reordering techniques on a variety of multi-threaded graph applications processing varied datasets.
       Our characterization reveals the inherent tension between reducing the cache footprint of hot vertices and preserving graph structure.

    \item We propose \dbg, a new skew-aware reordering technique that employs lightweight coarse-grain reordering to largely preserve existing graph structure while reducing the cache footprint of hot vertices. %

    \item Our evaluation on a real machine shows that \dbg \mbox{outperforms} existing skew-aware techniques.
    Averaging across 40 datapoints, \dbg yields a speed-up of 16.8\%, vs 11.6\% for the best-performing existing skew-aware \mbox{technique} over the baseline with no reordering.%

\squishend

\section{Background and Motivation\label{sec:background}}

\subsection{Properties of Natural Graphs\label{skew}}
Natural graphs have two distinguishing properties -- skew in their degree distribution and community structure.

\noindentsectiontitle{\bf Skew in Degree Distribution:}
Natural graphs exhibit \emph{skew} in their degree distribution~\cite{power-law,power-law-internet,powergraph,fc,hubcluster}.
The skew follows a power-law, with the vast majority of the vertices having relatively few edges and a small fraction of vertices featuring a large number of edges. Such skewed distribution is prevalent in many domains and found, for instance, in nodes in large-scale communication networks (e.g., the internet), web pages in web graphs, and individuals in social graphs.

Table~\ref{tab:skew-hot-vertices} quantifies the skew for the datasets evaluated in this work (Table~\ref{tab:datasets} in Sec.~\ref{sec:method} contains more details about these datasets).
For example, in the \sd dataset, 11\% of total vertices are classified as hot vertices in terms of their in-degree (13\% for out-degree) distribution. These hot vertices are connected to 88\% of all in-edges (88\% of all out-edges) in the graph. Similarly, in other datasets, 9\%-26\% of vertices are classified as hot vertices, which are connected to 80\%-94\% of all edges.

\noindentsectiontitle{\bf Community Structure:} 
Real-world graphs often feature \mbox{clusters} of highly interconnected vertices such as communities of \mbox{common} friends in a social graph~\cite{community1,community2}. 
Such community structure is often captured by vertex ordering within a graph dataset by placing vertices from the same community nearby in the memory space.
At runtime, vertices that are placed nearby in memory are typically processed within a short time window of each other. Thus, by placing vertices from the same community nearby in memory, both temporal and spatial locality is improved at the cache block level for such datasets.%

\subsection{Graph Processing Basics\label{graph-processing-basics}}
\begin{figure}
\centering
\subfloat{\label{csr-graph-a}{\transparent{1}\includegraphics[width=1px,height=1px]{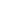}}}
\subfloat{\label{csr-graph-b}\includegraphics[width=0.99\linewidth]{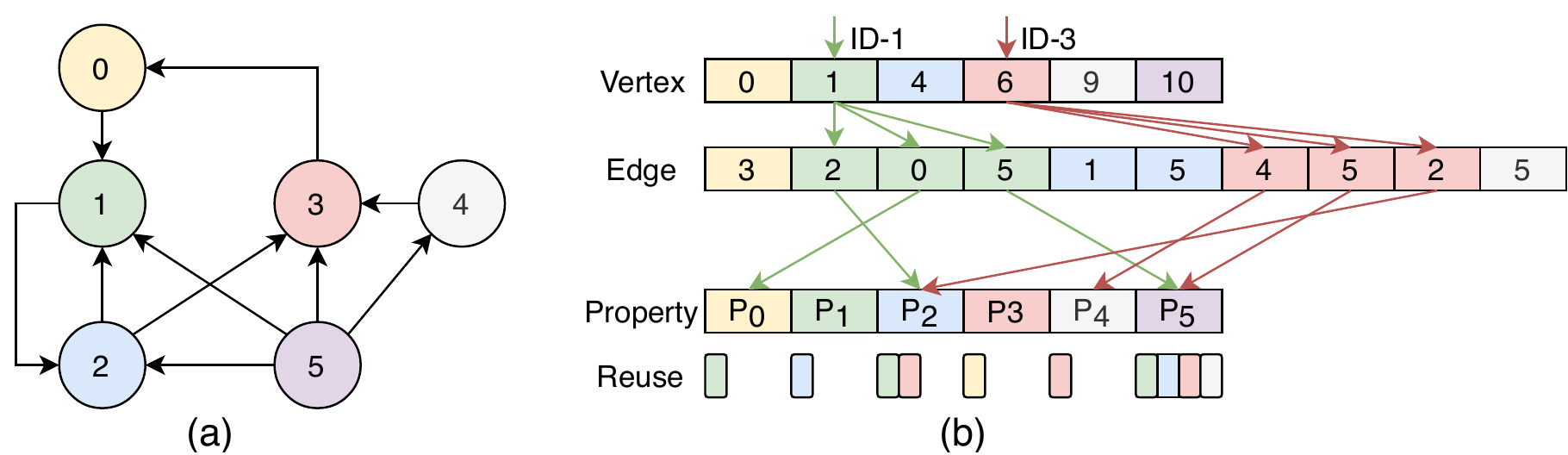}}
\subfloat{\label{csr-graph-c}{\transparent{1}\includegraphics[width=1px,height=1px]{pdfs/transparent.png}}}
\caption{(a) An example graph. (b) CSR format encoding in-edges. Elements of the same colors in all arrays, correspond to the same destination vertex. Number of bars (labeled \emph{Reuse}) below each element of the Property Array shows the number of times an element is accessed in one full iteration, where the color of a bar indicates the vertex making an access.\label{csr-graph}}
\end{figure}

\setlength\tabcolsep{4pt}
\begin{table}[!t]
    \centering
    \begin{tabularx}{1\linewidth}
        {|>{\centering\arraybackslash\hsize=0.15\hsize}X|
        >{\raggedleft\arraybackslash\hsize=0.45\hsize}X|
        >{\raggedleft\arraybackslash\hsize=0.05\hsize}X|
        >{\raggedleft\arraybackslash\hsize=0.05\hsize}X|
        >{\raggedleft\arraybackslash\hsize=0.05\hsize}X|
        >{\raggedleft\arraybackslash\hsize=0.05\hsize}X|
        >{\raggedleft\arraybackslash\hsize=0.05\hsize}X|
        >{\raggedleft\arraybackslash\hsize=0.05\hsize}X|
        >{\raggedleft\arraybackslash\hsize=0.05\hsize}X|
        >{\raggedleft\arraybackslash\hsize=0.05\hsize}X|
        }
    \hline
        & & \kr & \pl & \tw & \sd & \lj & \wl & \fr & \mpi \\ \hline \hline
        In & Hot Vertices (\%) & 9 &  16 &  12 &  11 &  25 &  12 &  24 &  10 \\
        Edges & Edge Coverage (\%) & 93 & 83 &  84 &  88 &  81 &  88 &  86 &  80 \\ \hline
        Out & Hot Vertices (\%) & 9 & 13 &  10 &  13 &  26 &  20 &  18 &  12 \\
        Edges & Edge Coverage (\%) & 93 & 88 &  83 &  88 & 82 & 94 & 92 & 81 \\ \hline
    \end{tabularx}
    \caption{\label{tab:skew-hot-vertices}Rows \#2 and \#4 show number of hot vertices (vertices with degree equal or greater than the average degree) as a percentage of total vertices, with respect to in-edges and out-edges, respectively; the higher the skew, the lower the percentage. Rows \#3 and \#5 show number of in-edges and out-edges connected to the hot vertices as a percentage of total edges, respectively; the higher the skew, the higher the percentage.
    }
\end{table}

The majority of shared-memory graph processing frameworks are written using a vertex-centric model, wherein an application computes some information for each vertex based on the properties of its neighboring vertices~\cite{ligra, galois, gap, graphmat, graphlab, graphchi}. Applications may perform pull- or push-based computations, or both. 
In pull-based computations, a vertex pulls (reads) property values of its in-neighbors (vertices whose edges point to this vertex). In push-based computations, a vertex pushes (writes) its property values to its out-neighbors (vertices pointed to by the edges of this vertex). This process may be iterative, and all or only a subset of vertices may participate in a given iteration.

The \emph{Compressed Sparse Row (CSR)} format is commonly used to represent graphs in a storage-efficient manner. 
CSR encodes in-edges for pull-based computations and out-edges for push-based computations. 
In this discussion, we focus on pull-based computations and note that the observations also hold for push-based computations. CSR uses a pair of arrays, \emph{Vertex} and \emph{Edge}, to encode the graph.
For every vertex, the Vertex Array maintains an index that points to its first in-edge in the Edge Array.
The Edge Array stores all in-edges, grouped by destination vertex ID. 
For each in-edge, the Edge Array entry stores the associated source vertex ID.%

Graph applications additionally use one or more \emph{Property Arrays} to hold partial or final results for every vertex. For example, the \emph{Pagerank} application maintains two rank values for every vertex; one computed from the previous iteration and one being computed in the current iteration. 
Implementation may use either two separate arrays (each storing one rank per vertex) or may use only one array (storing two ranks per vertex).
Fig.~\stitchref{csr-graph}{csr-graph-a} shows a simple graph and Fig.~\stitchref{csr-graph}{csr-graph-b} shows its CSR representation for pull-based computations along with one Property Array.

\subsection{Cache Behavior in Graph Analytics\label{sec:cache-behavior-in-graph-analytics}}

At the most fundamental level, a graph application computes a property for a vertex based on the properties of its neighbors.
To find the neighboring vertices, an application traverses the portion of the Edge Array corresponding to a given vertex, and then accesses elements of the Property Array corresponding to these neighboring vertices. Fig.~\stitchref{csr-graph}{csr-graph-b} shows the elements accessed while processing vertex ID-1 and ID-3. 

As the figure shows, each element of the Vertex and Edge Arrays is accessed exactly once during an iteration, exhibiting no temporal locality at LLC.
In contrast, the Property Array does exhibit temporal reuse. However, reuse is not consistent for all elements, being proportional to the number of out-edges for pull-based computations. Thus, %
high out-degree vertices exhibit high reuse. Fig.~\stitchref{csr-graph}{csr-graph-b} shows the reuse for high out-degree (\ie hot) vertices P$_2$ and P$_5$ of the Property Array assuming pull-based computations; the other elements do not exhibit reuse. The same observations apply to high in-degree vertices for push-based computations.

\setlength\tabcolsep{6pt}
\begin{table}[t]
    \centering
    \begin{tabularx}{1\linewidth}
        {|>{\centering\arraybackslash\hsize=0.2\hsize}X|
        >{\centering\arraybackslash\hsize=0.1\hsize}X|
        >{\centering\arraybackslash\hsize=0.1\hsize}X|
        >{\centering\arraybackslash\hsize=0.1\hsize}X|
        >{\centering\arraybackslash\hsize=0.1\hsize}X|
        >{\centering\arraybackslash\hsize=0.1\hsize}X|
        >{\centering\arraybackslash\hsize=0.1\hsize}X|
        >{\centering\arraybackslash\hsize=0.1\hsize}X|
        >{\centering\arraybackslash\hsize=0.1\hsize}X|
        }
    \hline
        Dataset & \kr & \pl & \tw & \sd & \lj & \wl & \fr & \mpi \\ \hline
        Avg.    & 1.3 & 1.6 & 1.5 & 1.8 & 3.5 & 3.1 & 2.7 & 2.6 \\
     \hline
    \end{tabularx}
    \caption{\label{tab:avg-hot-vertex}Average number of hot vertices per cache block. 
    \mbox{Calculation} assumes 8 bytes per vertex and 64 bytes per cache block, and counts only cache blocks that contain at least one hot vertex. As a result, any cache block can contain between 1--8 hot vertices. %
    }
\end{table}

\subsection{Poor Cache Efficiency for the Property Array\label{sec:challenges-property-array}}

As discussed in the previous section, the Property Array exhibits reuse and thus, should be the prime target for caching. However, not all elements of the array exhibit high reuse. Elements associated with hot vertices are the ones responsible for large amount of reuse. Despite such high reuse, on-chip cache capacity is severely underutilized as hot vertices are sparsely distributed throughout the memory space.

A cache block is typically comprised of multiple vertices as the properties associated with a vertex are much smaller than the size of a cache block. Thus, inevitably, hot vertices share space in a cache block with cold vertices, which poses a challenge for cache efficiency.
Even when such a cache block is retained in the cache, it leads to underutilization of cache capacity as a considerable fraction of the cache block is occupied by cold vertices that exhibit low or no reuse.  

Table~\ref{tab:avg-hot-vertex} shows the average number of hot vertices per cache block, assuming typical values of 8 bytes per vertex and 64 bytes per cache block. While, at best, 8 hot vertices can be packed together in a cache block, in practice, only 1.3 to 3.5 hot vertices are found per cache block across the datasets. As the footprint (\ie number of cache blocks) to store hot vertices is inversely proportional to the average number of hot vertices per cache block, the data shows significant opportunity in reducing the cache footprint of hot vertices, and in turn, improving cache efficiency.

\subsection{Vertex Reordering to Improve Cache Efficiency}

The order of vertices in memory is under the control of a graph application. Thus, the application can reorder vertices in memory before processing a graph to improve cache locality. To accomplish this, researchers have proposed various reordering techniques~\cite{gorder, rabbit, recall, LDG, METIS, SlashBurn, rcm, CHDFS, fc, hubcluster}. Reordering techniques only relabel vertices (and edges), which does not alter the graph itself and does not require any changes to the graph algorithms.
Following the relabeling, vertices (and edges) are reordered in memory based on the new vertex IDs. 

The most powerful reordering techniques like \gorder~\cite{gorder} leverage community structure, typically found in real-world graphs, to improve spatio-temporal locality. \gorder{} \mbox{comprehensively} analyzes the vertex connectivity and reorders vertices such that vertices that share common neighbors, and thus are likely to belong to the same community, are placed nearby in memory. While \gorder is effective at improving application performance, it requires a staggering reordering time that is often multiple orders of magnitude higher than the application runtime, rendering \gorder impractical~\cite{hubcluster}.

To keep reordering cost affordable, recent works have argued for skew-aware reordering techniques that reorder vertices solely based on vertex degrees~\cite{fc, hubcluster}.
By \mbox{relying} on lightweight analysis, these techniques can speed-up \mbox{applications} even after accounting for the reordering time. 

Skew-aware reordering techniques seek to reduce the cache footprint of hot vertices to improve cache efficiency. 
However, as a side-effect of reordering, they may destroy a graph's community structure, which could negate the performance gains achieved from the reduced footprint of hot vertices. Thus, there exists a tension between reducing the cache footprint of hot vertices and preserving graph structure when reordering vertices. In this context, we next detail existing skew-aware reordering techniques. 

\section{Skew-Aware Reordering Techniques\label{sec:lwr}}

\subsection{Objectives for High Performance Reordering\label{sec:objectives}}

In order to provide high performance for graph applications, skew-aware reordering techniques should achieve all of the following three objectives:

\visiblespace

\noindentsectiontitle{\bf O1. Low Reordering Time:} Reordering time plays a crucial role in deciding whether a technique is viable in providing end-to-end application performance after accounting for the reordering time. Lower reordering time facilitates amortizing the reordering overhead in a fewer graph traversals.

\noindentsectiontitle{\bf O2. High Cache Efficiency:} As explained in Sec.~\ref{sec:challenges-property-array}, a cache block is comprised of multiple vertices. Problematically, hot vertices are sparsely distributed throughout the memory space, which leads to cache blocks containing vertices with vastly different degrees. %
To address this, vertex reordering should ensure that hot vertices are placed adjacent to each other in the memory space, thus reducing the cache footprint of hot vertices, and in turn, improving cache efficiency.

\noindentsectiontitle{\bf O3. Structure Preservation:} As explained in Sec.~\ref{skew}, many real-world graph datasets have vertex ordering that results in high spatio-temporal cache locality. For such datasets, vertex reordering should ensure that the original structure is preserved as much as possible. If structure is not preserved, reordering may adversely affect the locality, negating performance gains achieved from the reduced footprint of hot vertices.

\subsection{Implications of Not Preserving Graph Structure\label{sec:random}}

\begin{figure}[!t]
    \centering
    \subfloat{\label{fig:reordering-existing-example-a}{\transparent{1}\includegraphics[width=1px,height=1px]{pdfs/transparent.png}}}
    \subfloat{\label{fig:reordering-existing-example-b}\includegraphics[width=0.995\linewidth]{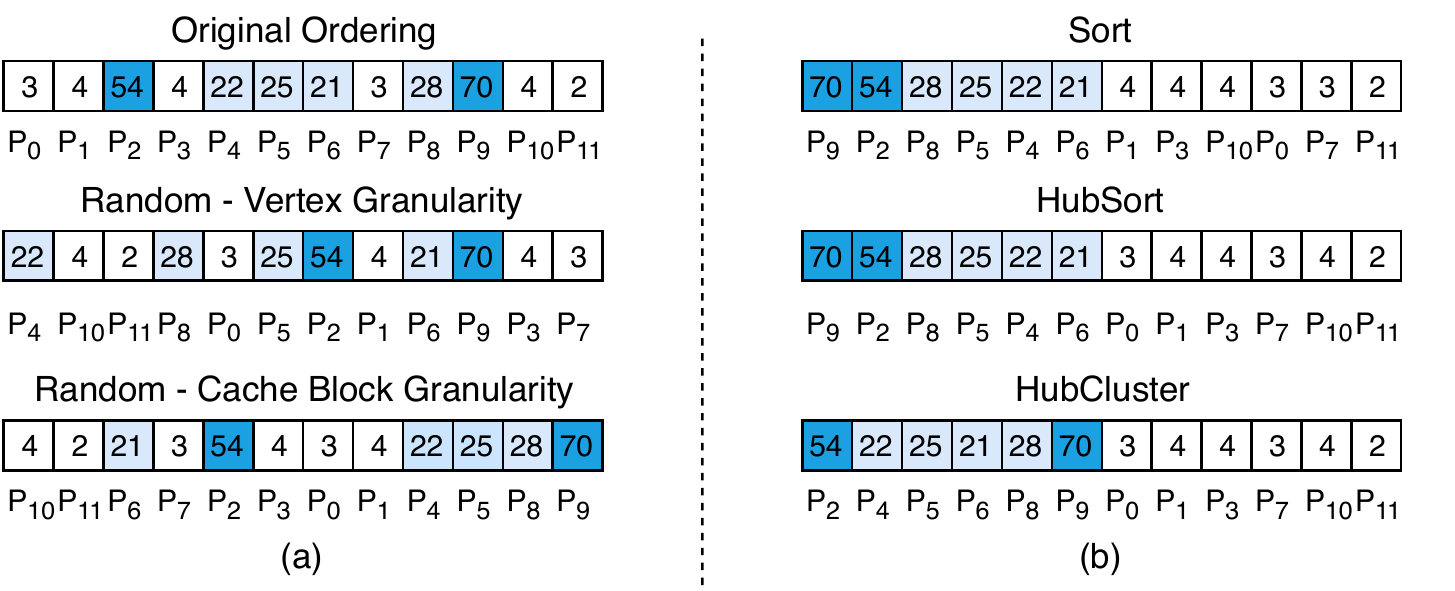}}
    \caption{Vertex ordering in memory for different techniques.
    Vertex degree is shown inside the box while original vertex ID is shown below the box.
    Hot vertices (degree $\ge$ 20) are shown in color. Hottest among the hot vertices (degree $\ge$ 40) are shown in darker shade. Finally, Random (Cache Block Granularity) assumes two vertices per cache block.}
    \label{fig:reordering-existing-example}
\end{figure}

In this section, we characterize how important it is to preserve graph structure for different datasets.
To \mbox{quantify} potential performance loss due to reduction in spatio-temporal locality arising from reordering, we randomly reorder vertices, which decimates any existing structure.
Randomly \mbox{reordering} all vertices would cause a slowdown for two potential \mbox{reasons}: 
(1) By destroying graph structure, thus reducing spatio-temporal locality.
(2) By further scattering hot vertices in memory, thus \mbox{increasing} the cache footprint of hot vertices.
To isolate performance loss due to the former, we also evaluate random reordering at a cache block granularity. In such a reordering, cache blocks (not individual vertices) are randomly reordered in memory, which means that the vertices within a cache block are moved as a group. As a result, the cache footprint of hot vertices is unaffected, and any change in performance can be directly attributed to a change in graph structure.
Fig.~\stitchref{fig:reordering-existing-example}{fig:reordering-existing-example-a} illustrates vertex placement in memory after Random Reordering at a vertex and at a cache block granularity.

\begin{figure}[t]
    \centering
    \includegraphics[width=1\linewidth]{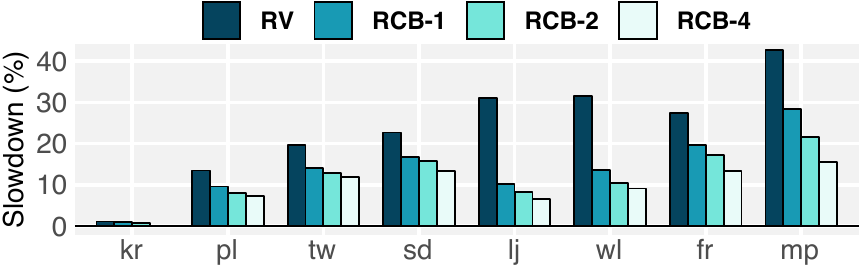}
    \caption{Application slowdown after random reordering at different granularity for the Radii application. The lower the bar, the better the application performance.\label{fig:random}}
\end{figure}

Fig.~\ref{fig:random} shows performance slowdown for Random Reordering for the Radii application on all datasets listed in Table~\ref{tab:datasets}.
The figure shows four configurations -- Random Vertex (RV) that reorders at a granularity of one vertex and Random Cache Block-$n$ (RCB-$n$) that reorders at a granularity of $n$ cache blocks, where $n$ is 1, 2 or 4.

Performance difference between RV and RCB-1 is very large for the four right-most datasets. Recall from Table~\ref{tab:avg-hot-vertex} that these datasets have relatively high number of hot vertices per cache block. RV scatters the hot vertices in memory, incurring large slowdowns for these datasets.

Performance slowdown for RCB-1 is significant on all real-world datasets (\ie all but \kr), and ranges from 9.6\% to 28.5\%.
This slowdown can be attributed to disruption in spatio-temporal locality for the real-world datasets, confirming existence of community structure in the original ordering of the datasets. 
As reordering granularity increases, disruption in graph structure reduces, which also reduces the slowdown.
For example, on the \mpi dataset, most affected by the Random Reordering, performance slowdown is 28.5\% for RCB-1, which 
reduces to 21.6\% for RCB-2 and 15.6\% for RCB-4.

Results for \kr{}, the only synthetic dataset in the mix, are in stark contrast with that of the real-world datasets.
As \kr{} is generated synthetically, \kr does not have any structure in the original ordering.
Thus, the performance on the \kr{} dataset is largely oblivious to random reordering at any granularity. 

The results show that the real-world graph datasets exhibit some structure in their original ordering, which, if not preserved, is likely to adversely affect the performance.
The results also indicate that structure can be largely preserved by applying reordering at a coarse granularity.

\subsection{Limitations of Skew-Aware Reordering Techniques\label{sec:existing-skew-aware}}

This section describes the existing skew-aware techniques and how they fare in achieving the three objectives listed in Sec.~\ref{sec:objectives}. As skew-aware techniques solely rely on vertex degrees for reordering, they all incur relatively low reordering time, achieving objective O1. However, for the two remaining objectives -- reducing the cache footprint of hot vertices and preserving existing graph structure --  existing techniques trade one for the other, hence failing to achieve at
least one of the two objectives. 

\setlength\tabcolsep{3pt}
\begin{table}[t]
    \centering
    \begin{tabularx}{1\linewidth}
        {|>{\centering\arraybackslash\hsize=0.4\hsize}X|
          >{\raggedleft\arraybackslash\hsize=0.075\hsize}X|
          >{\raggedleft\arraybackslash\hsize=0.075\hsize}X|
          >{\raggedleft\arraybackslash\hsize=0.075\hsize}X|
          >{\raggedleft\arraybackslash\hsize=0.075\hsize}X|
          >{\raggedleft\arraybackslash\hsize=0.075\hsize}X|
          >{\raggedleft\arraybackslash\hsize=0.075\hsize}X|
          >{\raggedleft\arraybackslash\hsize=0.075\hsize}X|
          >{\raggedleft\arraybackslash\hsize=0.075\hsize}X|
          }
    \hline
        Per-Vertex Property & \kr & \pl & \tw & \sd & \lj & \wl & \fr & \mpi \\ \hline \hline
        8 Bytes & 44 & 51 & 56 & 80 & 9 &  16 & 115 & 39 \\ \hline
        16 Bytes & 88 & 102 & 112 & 160 & 18 & 32 & 230 & 78 \\ \hline
    \end{tabularx}
    \caption{\label{tab:fp-hot-vertices} Cache size in MB needed to store {\em all hot vertices}, for 8 and 16 bytes per property, respectively. Vertex is classified hot if its degree is equal or greater than the average degree of the dataset.}  
\end{table}

\setlength\tabcolsep{1pt}
\begin{table}[t]
    \centering
    \begin{tabularx}{1\linewidth}
        {|>{\centering\arraybackslash\hsize=0.22\hsize}X|
          >{\raggedleft\arraybackslash\hsize=0.12\hsize}X|
          >{\raggedleft\arraybackslash\hsize=0.12\hsize}X|
          >{\raggedleft\arraybackslash\hsize=0.125\hsize}X|
          >{\raggedleft\arraybackslash\hsize=0.135\hsize}X|
          >{\raggedleft\arraybackslash\hsize=0.145\hsize}X|
          >{\raggedleft\arraybackslash\hsize=0.135\hsize}X|
          }
        \hline 
        Degree Range & {\footnotesize[1\avgdegree,2\avgdegree)} & {\footnotesize[2\avgdegree,4\avgdegree)} & {\footnotesize[4\avgdegree,8\avgdegree)} & {\footnotesize[8\avgdegree,16\avgdegree)} & {\footnotesize[16\avgdegree,32\avgdegree)} & {\footnotesize[32\avgdegree,\infinity)} \\ \hline
        Vertices (\%) & 45\% & 28\% & 15\% & 7\% & 3\% & 2\%  \\ \hline
        Footprint & 35.8 & 22.3 & 12.0 & 5.7 & 2.2 & 1.8 \\ \hline  
    \end{tabularx}
    \caption{\label{tab:degree-range}
    Degree distribution of hot vertices for the \sd dataset, whose Average Degree (\avgdegree) is 20. Row \#2 shows percentage of total hot vertices while row \#3 shows the footprint requirement in MB.
    }
\end{table}
\visiblespace

\noindentsectiontitle{\bf Sort} reorders vertices based on the descending order of their degree.
Sort requires the least possible number of cache blocks to store hot vertices without explicitly classifying individual vertices as hot or cold. 
However, sort reorders \emph{all} vertices, which completely destroys the original graph structure. 
Fig.~\stitchref{fig:reordering-existing-example}{fig:reordering-existing-example-b} shows vertex placement in memory after the Sort Reordering.

\noindentsectiontitle{\bf Hub Sorting}~\cite{fc} (also known as Frequency-based Clustering) was proposed as a variant of Sort that aims to preserve some structure while reducing the cache footprint of hot vertices.
Hub Sorting uses a threshold (typically, average degree of the dataset) to classify vertices as hot or cold, and only sorts the hot vertices.%

Hub Sorting does preserve partial structure by not sorting the cold vertices, but problematically, the hot vertices are fully sorted. While hot vertices constitute a smaller fraction compared to the cold ones, recall from Table~\ref{tab:skew-hot-vertices} that 
hot vertices account for up to 26\% of the total vertices. Moreover, hot vertices are connected to the vast majority of edges (80\%-94\%) and thus, are responsible for the majority of reuse. Consequently, preserving structure for hot vertices is also important, at which Hub Sorting fails.

\noindentsectiontitle{\bf Hub Clustering}~\cite{hubcluster} is a variant of Hub Sorting that only segregates hot vertices from the cold ones but does not sort them. While Hub Clustering was proposed as an alternative to Hub Sorting that has lower reordering time, we note that Hub Clustering is also better than Hub Sorting at preserving existing graph structure as Hub Clustering does not sort any vertices. However, by not sorting hot vertices, Hub Clustering sacrifices significant opportunity in
improving cache efficiency as discussed next.

For large graph datasets, it is unlikely that all hot vertices fit in the LLC. For example, the \sd dataset requires at least 80MB to store all hot vertices assuming only 8 bytes per vertex (refer to Table~\ref{tab:fp-hot-vertices} for requirements of the remaining datasets). The required capacity significantly exceeds typical LLC size of commodity server processors. As a result, all hot vertices compete for the
limited LLC capacity, causing cache thrashing.

Fortunately, not all hot vertices have similar reuse, as vertex degree varies vastly among hot vertices. 
Table~\ref{tab:degree-range} shows the degree distribution for just the hot vertices of the \sd dataset. 
Each column in the table represents a degree range as a function of \avgdegree, the average degree of the dataset. For instance, the first column covers vertices whose degree ranges from \avgdegree to 2\avgdegree; these are the lowest-degree vertices among the hot ones (recall that a vertex is classified as hot if its degree is equal or greater than \avgdegree). 
For a given range, the table shows number of vertices (as a percentage of total hot vertices) whose degree is within that range. The table also shows cache capacity needed for those many vertices assuming 8 bytes per vertex property.
Unsurprisingly, given the power-law degree distribution, the table shows that the least-hot vertices are the most numerous, representing 45\% of all hot vertices and requiring 35.8MB capacity, yet likely exhibiting the least reuse among hot vertices. 
In contrast, vertices with degree above 8\avgdegree (three right-most columns) are the hottest of all, constituting just 12\% of total hot vertices ($<10$MB footprint).
Naturally, these hottest vertices are the ones that should be retained in the cache.
However, by not sorting hot vertices, Hub Cluster fails to differentiate between the most- and the least-hot vertices, hence denying the hottest vertices an opportunity to stay in the cache in the presence of cache thrashing. 

\visiblespacel

To summarize, Sort achieves the maximum reduction in the cache footprint of hot vertices. However, in doing so, Sort completely decimates existing graph structure. Hub Sorting and Hub Clustering both classify vertices as hot or cold based on their degree and preserve the structure for cold vertices. However, in dealing with hot vertices, they resort to inefficient extremes. 
At one extreme, Hub Sorting employs fine-grain reordering that sorts all hot vertices, destroying existing graph structure.
At the other extreme, Hub Clustering does not apply any kind of reordering among hot vertices, sacrificing significant opportunity in improving cache efficiency. %

\section{Degree-Based Grouping (DBG)\label{sec:spot}}
To address the limitations of existing skew-aware reordering techniques, we propose \emph{\dbgfullname (\dbg)}, a novel skew-aware technique that applies coarse-grain reordering such that each cache block is exclusively comprised of vertices having similar degree, and in turn, similar hotness, while also preserving graph structure at large.

Unlike Hub Sorting and Hub Clustering, which rely on a single threshold to classify vertices as hot or cold, \dbg employs a simple binning algorithm to \emph{coarsely} partition vertices into different groups (or bins) based on their hotness level.
Groups are assigned exclusive degree ranges such that the degree of any vertex falls within a degree range of exactly one group.
Within each group, \dbg maintains the original relative order of vertices to preserve graph structure at large.
To keep the reordering time low, \dbg maintains only a small number of groups and does not sort vertices within any group. 
Listing~\ref{spot-listing} presents the formal \dbg algorithm.

\begin{lstlisting}[caption=\dbg algorithm. Degree can be in-degree or out-degree or sum of both., label=spot-listing, float, floatplacement=t, aboveskip=0.65em, morekeywords={D,M,K,P,Q,G,V,E,Group},basicstyle=\small,captionpos=b]
G(V, E) where Graph G has V vertices and E edges.
%*\emph{Input}*): Degree Distribution D[], where D[v] is degree of vertex v.
%*\emph{Output}*): Mapping M[], where M[v] is the new ID of vertex v.
%*\emph{\dbg}*): Binning algorithm to reorder vertices into K groups (K $>$ 0). 

1: Assign contiguous range [P$_k$, Q$_k$) to every Group$_k$ such that,
       Q$_1$ > max(D[])    &    
       P$_K$ $\leq$ min(D[])    &
       Q$_{k+1}$ = P$_k$  < Q$_k$,    for every k < K
       
2: For every vertex v from 1 to V
       Append v to the Group$_k$ for which D[v] $\in$ [P$_k$, Q$_k$).
       
3: Assign new IDs to all vertices as follows:
       id := 1
       For every Group$_k$ from 1 to K
            For every vertex v in Group$_k$
                M[v] := id++, where v is the original ID
\end{lstlisting}

\begin{figure}[!t]
    \centering
    \includegraphics[width=1\linewidth]{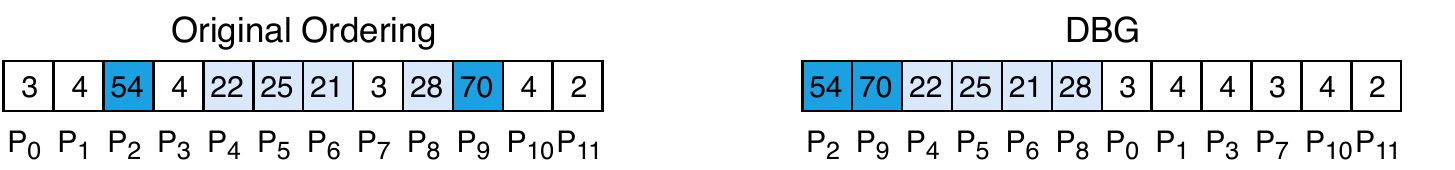}
    \caption{Vertex ordering in memory after DBG. In this example, \dbg partitions vertices into three groups with degree ranges [0, 20), [20, 40) and [40, 80). \dbg maintains a relative order of vertices within a group. As a result, many vertices are placed nearby the same vertices as before the reordering such as vertex sets (P$_4$, P$_5$, P$_6$), (P$_0$, P$_1$) and (P$_{10}$, P$_{11}$).}
    \label{fig:reordering-example}
\end{figure}

To assign degree ranges to different groups, \dbg leverages the power-law distribution of vertex connectivity in natural graphs. %
For example, recall Table~\ref{tab:degree-range}, which shows distribution of vertices across different degree ranges. 
Vertices with the smallest degree range constitute the largest fraction of hot vertices. As degree range doubles, the number of vertices are roughly halved, exhibiting the power-law distribution.
Thus, geometrically-spaced degree ranges provide a natural way to segregate vertices with different levels of hotness. 
At the same time, using such wide ranges to partition vertices facilitates reordering at a very coarse granularity, preserving structure at large. 
Meanwhile, by not sorting vertices within any group, \dbg incurs a very low reordering time.  
Thus, \dbg successfully achieves all three objectives listed in Sec.~\ref{sec:objectives}.
Fig.~\ref{fig:reordering-example} shows vertex placement in memory after the \dbg Reordering, for a synthetic example. 

Finally, we note that the \dbg algorithm (Listing~\ref{spot-listing}) provides a general framework to understand trade-offs between reducing the cache footprint of hot vertices and preserving graph structure just by varying a number of groups and their degree ranges. Indeed, Table~\ref{tab:spot-bin} shows how different skew-aware techniques can be implemented using the \dbg algorithm. 
For example,
Hub Clustering can be viewed as an implementation of \dbg algorithm with two groups -- one containing hot vertices and another one containing cold vertices.
Similarly,
Sort can be seen as an implementation of \dbg algorithm with as many number of groups as many unique
degrees a given dataset has. 
Consequently, for a given unique degree, the associated group contains all vertices having the same degree, effectively sorting vertices by their degree.
In general, as the number of groups is increased, the degree range gets narrower and vertex reordering gets finer, causing more disruption to existing structure.
Table~\ref{tabl:spot-cost} qualitatively compares \dbg to prior techniques.

\setlength\tabcolsep{2pt}
\begin{table}[!t]
    \centering
    \begin{tabularx}{1\linewidth}
        {|>{\raggedright\arraybackslash\hsize=0.17\hsize}X| %
          >{\centering\arraybackslash\hsize=0.2\hsize}X|
          >{\raggedright\arraybackslash\hsize=0.63\hsize}X|
          }
        \hline
        \centering Reordering & \#groups & \centering Degree Range \tabularnewline \hline
        \hline
        Sort & \maxdegree{}$+1$ & [$n$, $n+1$) where n $\in$ [0, \maxdegree{}] \\ \hline
        Hub     & \multirow{2}{*}{\maxdegree{}-\avgdegree{}$+2$} & [0, \avgdegree), \\ 
        Sorting                                &                                   & [$n$, $n+1$) where n $\in$ [\avgdegree{}, \maxdegree{}] \\ \hline 
        Hub & \multirow{2}{*}{2} & [0, \avgdegree{}), \\
        Clustering & & [\avgdegree{}, \maxdegree{}] \\ \hline
        \multirow{2}{*}{\dbg} & \multirow{2}{*}{$\floor{log_{2}{\frac{\mathbb{M}}{\mathbb{C}}}} + 2$} 
 & [0, \constant), \\
            &   & [2$^n$\constant, 2$^{n+1}$\constant) where n $\in$ \bigg[0, $\floor{log_{2}{\frac{\mathbb{M}}{\mathbb{C}}}}\bigg]$ \\ \hline
    \end{tabularx}
    \caption{Implementation of various skew-aware techniques using \dbg algorithm. 
    \avgdegree is the average and \maxdegree is the maximum degree of the dataset.
    For \dbg, \constant is some threshold such that 0 $<$ \constant $<$ \maxdegree. 
    \label{tab:spot-bin}}
\end{table}

\setlength\tabcolsep{4pt}
\begin{table}[!t]
    \centering
    \begin{tabularx}{1\linewidth}
        {|>{\raggedright\arraybackslash\hsize=0.4\hsize}X| %
          >{\centering\arraybackslash\hsize=0.215\hsize}X|
          >{\centering\arraybackslash\hsize=0.2\hsize}X|
          >{\centering\arraybackslash\hsize=0.22\hsize}X|
        }
    \hline
    \multirow{2}{*}{Technique} &
    \multicolumn{1}{c|}{Structure} &
    \multicolumn{1}{c|}{Reordering} &
    \multicolumn{1}{c|}{Net} \\
    
    &
    \multicolumn{1}{c|}{Preservation} &
    \multicolumn{1}{c|}{Time} &
    \multicolumn{1}{c|}{Performance} \\ \hline
    \hline
       
    Sort &
    \xmark &
    \cmark &
    \cmark \\ \hline
    
    Hub Sorting~\cite{fc} &
    \cmark &
    \cmark &
    \cmark \\ \hline
    
    Hub Clustering~\cite{hubcluster} &
    \cmark\cmark &
    \cmark\cmark &
    \cmark  \\ \hline
    
    \rowcolor{lightgray}
    \dbg (proposed) &
    \cmark\cmark &
    \cmark\cmark &
    \cmark\cmark \\ \hline

    \gorder~\cite{gorder} &
    \cmark\cmark &
    \xmark &
    \xmark \\ \hline
 
    \end{tabularx}
    \caption{\label{tabl:spot-cost}Qualitative performance of different reordering techniques for graph analytics on natural graphs.}
\end{table}

\section{Methodology}
\label{sec:method}

\subsection{Graph Processing Framework, Applications and Datasets}
\label{sec:framework}
For the evaluation, we use \emph{Ligra}~\cite{ligra}, a widely used shared-memory graph processing framework that supports both pull- and push-based computations, including switching from pull to push (and vice versa) at the start of a new iteration. We evaluate various reordering techniques using five iterative graph applications listed in Table~\ref{tab:graph-worklads}, on eight graph datasets listed in Table~\ref{tab:datasets}, resulting in 40 datapoints for each technique.

We obtained the source code for the graph applications from Ligra~\cite{ligra}. Implementation of the graph applications is unchanged except for an addition of an array to keep a mapping between the vertex ID assignments before and after reordering. The mapping is needed to ensure that root-dependent traversal applications running on the reordered graph datasets use the same root as the baseline execution running on the original graph dataset. We compile the applications using g++−6.4 with O3
optimization level on Ubuntu 14.04.1 booted with Linux kernel 4.4.0-96-lowlatency and use OpenMP for parallelization. To utilize memory bandwidth from both sockets, we run every application under NUMA interleave memory allocation policy.
Table~\ref{tab:graph-workloads-properties} lists various properties for the Ligra implementation of the evaluated graph applications. 

\setlength\tabcolsep{4pt}
\begin{table}[!t]
    \centering
    \begin{tabularx}{1\linewidth}
        {|>{\centering\arraybackslash\hsize=0.24\hsize}X| 
          >{\raggedright\arraybackslash\hsize=0.76\hsize}X|
        }
        \hline
        Application    & \multicolumn{1}{c|}{Brief description} \\ \hline
        \hline
        {Betweenness Centrality (BC)} & finds the most central vertices in a graph by using a BFS kernel to count the number of shortest paths passing through each vertex from a given root vertex. \\ \hline
        {Single Source Shortest Path (SSSP)} & computes shortest distance for vertices in a weighted graph from a given root vertex using the Bellman Ford algorithm. \\ \hline
        {Pagerank ~~~~~~(PR)}  & is an iterative algorithm that calculates ranks of vertices based on the number and quality of incoming edges to them~\cite{pagerank}. \\ \hline
        {PageRank-Delta (PRD)} &is a faster variant of PageRank in which vertices are active in an iteration only if they have accumulated enough change in their PageRank score. \\ \hline %
        {Radii Estimation (Radii)} & estimates the radius of each vertex by performing multiple parallel BFS's from a small sample of vertices~\cite{radii}. \\ \hline
    \end{tabularx}
    \caption{\label{tab:graph-worklads}A list of graph applications evaluated in this work.}
\end{table}

\setlength\tabcolsep{2pt}
\begin{table}[!t]
    \centering
    \begin{tabularx}{1\linewidth}
        {|>{\raggedright\arraybackslash\hsize=0.12\hsize}X|
          >{\raggedright\arraybackslash\hsize=0.2\hsize}X| 
          >{\raggedleft\arraybackslash\hsize=0.2\hsize}X| 
          >{\raggedleft\arraybackslash\hsize=0.3\hsize}X| 
          >{\centering\arraybackslash\hsize=0.18\hsize}X|
        }
        \hline
        Graph & \centering{Computation}  & \multicolumn{2}{c|}{Per-Vertex Property Size (bytes)} & \centering{Degree}    \tabularnewline \cline{3-4}
        App. &\centering{Type} & \centering{All Properties} & \centering{Only Properties with Irregular Accesses} & \centering{Type used for Reordering} \tabularnewline \hline \hline
        BC & pull-push & 17 & 8 & out \tabularnewline \hline 
        SSSP & push-only & 8 & 8 & in \tabularnewline \hline 
        PR & pull-only & 20 & 12 & out \tabularnewline \hline 
        PRD & push-only & 20 & 8 & in \tabularnewline \hline 
        Radii & pull-push & 20 & 8 & out \tabularnewline \hline 
    \end{tabularx}
    \caption{Properties of graph applications. 
    In addition to the vertex properties, all graph applications require 4 bytes to encode a vertex and 8 bytes to encode an edge. 
    }
    \label{tab:graph-workloads-properties}
\end{table}

\subsection{Evaluation Platform and Methodology\label{sec:soft-eval}}

Evaluation is done on a dual-socket server with two Broadwell based {\em Intel Xeon CPU E5-2630}~\cite{xeon}, each with 10 cores clocked at 2.2GHz and a 25MB shared LLC. Hyper-threading is kept on, exposing 40 hardware execution contexts across both CPUs. Server has 128GB of DRAM provided by eight DIMMs clocked at 2133MHz. Applications use 40 threads, and the threads are pinned to avoid performance variations due to OS scheduling. To further reduce
sources of performance variation, DVFS features are disabled. Finally, \emph{Transparent Huge Pages} is kept on to reduce TLB misses.

\setlength\tabcolsep{1pt}
\begin{table}[!t]
    \centering
    \begin{tabularx}{1\linewidth}
    {|>{\raggedright\arraybackslash\hsize=0.33\hsize}X| 
      >{\raggedleft\arraybackslash\hsize=0.1\hsize}X|
      >{\raggedleft\arraybackslash\hsize=0.11\hsize}X| 
      >{\raggedleft\arraybackslash\hsize=0.11\hsize}X| 
      >{\centering\arraybackslash\hsize=0.14\hsize}X| 
      >{\centering\arraybackslash\hsize=0.21\hsize}X| 
    }
    \hline
        { \centering \multirow{2}{*}{Dataset} }                              &  \centering Vertex    & \centering Edge   & \centering Avg.& \centering \multirow{2}{*}{Type}  & \centering Original \tabularnewline
        {  }                                                &  \centering Count     & \centering Count  & \centering Degree &                   & \centering Ordering  \tabularnewline
    \hline
    \hline 
        {Kron ({\em kr})~\cite{gap}}                        & \num{67}{M}           & \num{1323}{M}     & 20 & Synthetic            & Unstructured                   \tabularnewline \hline
        {PLD ({\em pl})~\cite{pld}}                         & \num{43}{M}           & \num{623}{M}      & 15 & Real              & Unstructured          \tabularnewline \hline
        {Twitter ({\em tw})~\cite{twitter}}                 & \num{62}{M}           & \num{1468}{M}     & 24 & Real              & Unstructured          \tabularnewline \hline
        {SD ({\em sd})~\cite{pld}}                          & \num{95}{M}           & \num{1937}{M}     & 20 & Real              & Unstructured          \tabularnewline \hline
        {LiveJournal ({\em lj})~\cite{snapnets}}            & \num{5}{M}            & \num{68}{M}       & 14 & Real              & Structured            \tabularnewline \hline
        {WikiLinks ({\em wl})~\cite{konect-wl}}       & \num{18}{M}           & \num{172}{M}      & 9 & Real              & Structured            \tabularnewline \hline
        {Friendster ({\em fr}) \cite{konect-friendster}}    & \num{64}{M}           & \num{2147}{M}     & 33 & Real              & Structured            \tabularnewline \hline
        {MPI ({\em mp})~\cite{twitter_mpi}}                 & \num{53}{M}           & \num{1963}{M}     & 37 & Real              & Structured            \tabularnewline \hline
    \end{tabularx}
    \caption{\label{tab:datasets}Properties of the evaluated graph datasets. 
    We empirically label those datasets as {\em sturctured} on which Random Reordering (RV) causes more than 25\% slowdown (Fig.~\ref{fig:random}).
    }
\end{table}

\setlength\tabcolsep{2pt}
\begin{table}[!t]
    \centering
    \begin{tabularx}{1\linewidth}
    {|>{\raggedright\arraybackslash\hsize=0.52\hsize}X| 
      >{\raggedleft\arraybackslash\hsize=0.11\hsize}X|
      >{\raggedleft\arraybackslash\hsize=0.11\hsize}X| 
      >{\raggedleft\arraybackslash\hsize=0.11\hsize}X| 
      >{\centering\arraybackslash\hsize=0.15\hsize}X| 
    }
    \hline
        { \centering \multirow{2}{*}{Dataset} } &  \centering Vertex    & \centering Edge   & \centering Avg.& \centering \multirow{2}{*}{Type}  \tabularnewline
        &  \centering Count     & \centering Count  & \centering Degree & \tabularnewline
    \hline
    \hline 
        {Uniform ({\em uni})~\cite{PaRMAT}} & \num{50}{M} & \num{1000}{M} & 20.0 & Synthetic \tabularnewline \hline
        {USA Road Network ({\em road})~\cite{road}} & \num{24}{M} & \num{29}{M} & 1.2 & Real \tabularnewline \hline
    \end{tabularx}
    \caption{\label{tab:no-skew-datasets}Properties of the no-skew graph datasets. The {\em uni} dataset is generated using R-MAT~\cite{rmat} methodology with parameter values of A=B=C=25.
    }
\end{table}

We evaluate each reordering technique on every combination of graph applications and graph datasets 11 times, and record the average runtime of 10 executions, excluding the timing of the first execution to allow the caches to warm up. We report the speed-up over the entire application runtime (with and without reordering cost) but exclude the graph loading time from the disk. For iterative applications -- \pr{} and \prd{} -- we run them until convergence and consider the aggregate runtime over all
iterations. For root-dependent traversal applications -- \sssp{} and \bc{} -- we run them from eight different root vertices for each input dataset and consider the aggregate runtime over all eight traversals. Finally, we note that the application runtime is relatively stable across executions. For each reported datapoint, coefficient of variation is at most 2.3\% for PRD and at most 1.6\% for other applications.

\subsection{Reordering Techniques}

We evaluate \dbg and compare it with all three existing skew-aware techniques described in Sec.~\ref{sec:existing-skew-aware} (Sort, HubSort~\cite{fc} and HubCluster~\cite{hubcluster}) along with \gorder{}~\cite{gorder} -- the state-of-the-art structure-aware reordering technique.

For \gorder{}, we use the source code available from the authors. As \gorder is only available in a single-thread implementation, while reporting the reordering time of \gorder for a given dataset, we optimistically divide the reordering time by 40 (maximum number of threads supported on the server) to provide a fair comparison with skew-aware techniques whose reordering implementation is fully parallelized.

For \dbg{}, we use 8 groups with the ranges [32\avgdegree, \infinity), [16\avgdegree, 32\avgdegree), [8\avgdegree, 16\avgdegree), [4\avgdegree, 8\avgdegree), [2\avgdegree, 4\avgdegree), [1\avgdegree, 2\avgdegree), [\avgdegree/2, \avgdegree) and [0, \avgdegree/2), where \avgdegree{} is the average degree of the graph dataset. Note that we also partition cold vertices into two groups. We developed a multi-threaded implementation of \dbg, which is available at
\url{https://github.com/faldupriyank/dbg}.

\begin{figure}[!t]
    \centering
    \includegraphics[width=1\linewidth]{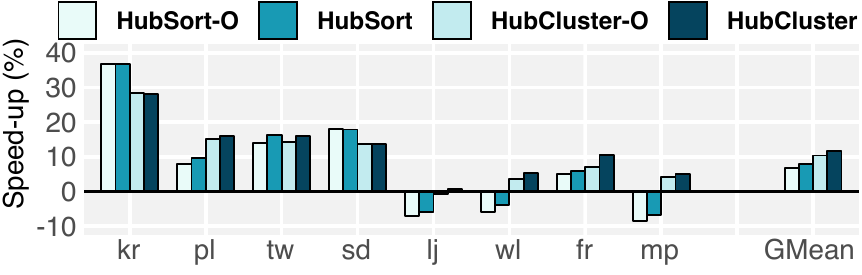}
    \caption{Application speed-up over the baseline with no reordering. 
    Techniques with suffix O use their original implementations whereas techniques without any suffix are implemented using \dbg algorithm as per Table~\ref{tab:spot-bin}.
    The bars for the datasets show geometric mean of speed-ups across five applications for a given dataset.
    }
    \label{fig:hub-s-perf}
\end{figure}

Finally, we implement HubSort and HubCluster using the \dbg algorithm as shown in Table~\ref{tab:spot-bin}. We found our implementations to be more effective than the original implementations (referred to as HubSort-O and \mbox{HubCluster-O}) provided by the authors of HubCluster.
Fig.~\ref{fig:hub-s-perf} shows application speed-up over the baseline with no reordering. Table~\ref{tab:reordering-time-norm} shows reordering time normalized to that of Sort.  As our implementation of both techniques provides better speed-ups and lower reordering time, we use our implementations in the main evaluation.
\begin{table}[t]
    \centering
    \begin{tabularx}{1\linewidth}
        {|>{\raggedright\arraybackslash\hsize=0.28\hsize}X| 
          >{\centering\arraybackslash\hsize=0.09\hsize}X|
          >{\centering\arraybackslash\hsize=0.09\hsize}X|
          >{\centering\arraybackslash\hsize=0.09\hsize}X|
          >{\centering\arraybackslash\hsize=0.09\hsize}X|
          >{\centering\arraybackslash\hsize=0.09\hsize}X|
          >{\centering\arraybackslash\hsize=0.09\hsize}X|
          >{\centering\arraybackslash\hsize=0.09\hsize}X|
          >{\centering\arraybackslash\hsize=0.09\hsize}X|
        }
        \hline
        Technique	    &kr	&pl	&tw	&sd	&lj	&wl	&fr	&mp \tabularnewline \hline \hline
        HubSort-O	        &1.02	&1.04	&1.01	&1.02	&1.09	&0.79	&1.04	&1.01 \tabularnewline 
        \hline \rowcolor{lightgray}
        HubSort	    &0.80	&0.82	&0.84	&0.84	&0.87	&0.91	&0.90	&0.89 \tabularnewline \hline \hline
        HubCluster-O	    &0.78	&0.79	&0.81	&0.81	&0.78	&0.56	&0.88	&0.87 \tabularnewline 
        \hline \rowcolor{lightgray}
        HubCluster	&0.77	&0.74	&0.81	&0.78	&0.76	&0.81	&0.84	&0.82 \tabularnewline \hline
    \end{tabularx}
    \caption{Reordering time for existing skew-aware techniques, normalized to that of Sort. Lower is better.}
    \label{tab:reordering-time-norm}
\end{table}

\section{Evaluation\label{sec:eval}}

\begin{figure*}[!t]
    \centering
    \subfloat[\em Unstructured datasets.]{\label{fig:all-app-perf-unstructured}\includegraphics[width=1\linewidth]{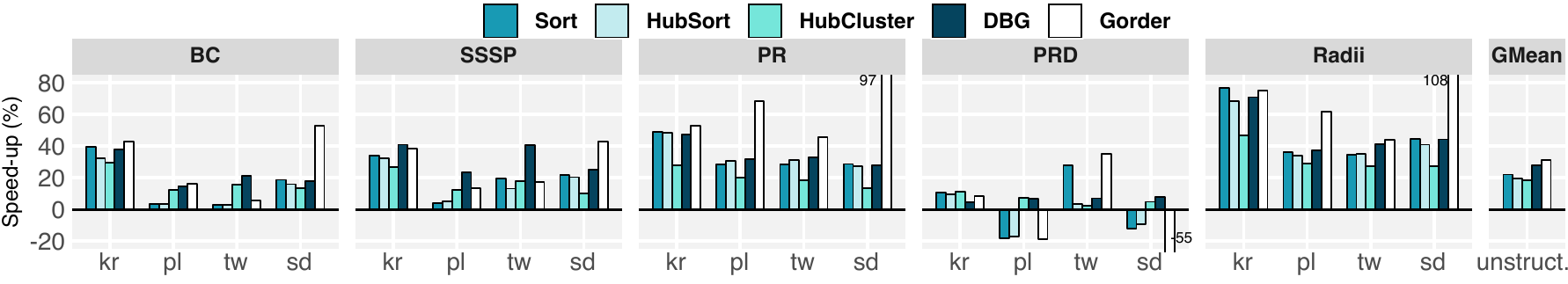}}\\
    \subfloat[\em Structured datasets.]{\label{fig:all-app-perf-structured}\includegraphics[width=1\linewidth]{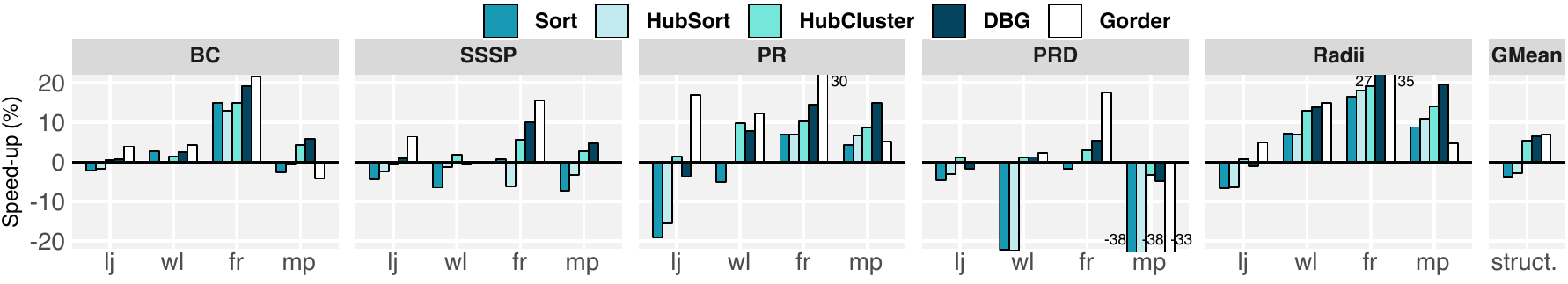}}
    \caption{\label{fig:all-app-perf}Application speed-up (excluding reordering time) for reordering techniques over the baseline with no reordering.}
\end{figure*}

In this section, we evaluate the effectiveness of \dbg against the state-of-the-art reordering techniques. In Sec.~\ref{sec:sw-excluding-cost}, we compare the application speed-up for these techniques {\em without considering the reordering time}. In Sec.~\ref{sec:pull-dominated} and Sec.~\ref{sec:push-dominated}, we  analyze different levels of cache hierarchy to understand the sources of performance variation. Subsequently, to understand  the effect of the reordering time on end-to-end performance, we
compare the application speed-up {\em after accounting for the reordering time} in Sec.~\ref{sec:sw-including-cost}.

\subsection{Performance Excluding Reordering Time\label{sec:sw-excluding-cost}}

Fig.~\ref{fig:all-app-perf} shows application speed-up excluding reordering time for various datasets. Averaging across all 40 datapoints (combining all structured and unstructured), \dbg provides 16.8\% speed-up over the baseline with no reordering, outperforming all existing skew-aware techniques: Sort (8.4\%), HubSort (7.9\%) and HubCluster (11.6\%). \gorder, which comprehensively analyzes graph structure, yields 18.6\% average speed-up, marginally higher than that of \dbg. We
next analyze performance variations across datasets and applications.

\subsubsection{\textbf{\textit{Unstructured vs Structured}}}
As shown in Fig.~\stitchref{fig:all-app-perf}{fig:all-app-perf-unstructured}, on unstructured datasets, all reordering techniques provide positive speed-ups for all applications except for PRD.
Sec.~\ref{sec:push-dominated} explains the reasons for slowdowns for the PRD application.
Among skew-aware techniques, \dbg provides the highest average speed-up of 28.1\% in comparison to 22.1\% for Sort, 19.8\% for HubSort and 18.3\% for HubCluster.

On synthetic dataset \kr, all techniques except HubCluster provide similar speed-ups as \kr is largely insensitive to structure preservation.
Similarly, on other unstructured datasets, as hot vertices are relatively more scattered in memory (see Table~\ref{tab:avg-hot-vertex}), the benefit of vertex packing outweighs potential slowdown due to structure disruption. Thus, Sort, despite completely decimating the original graph structure, outperforms HubSort and HubCluster on more than half of the datapoints.
Meanwhile, \dbg, which also preserves graph structure while reducing the cache footprint of hot vertices, provides higher performance than Sort on more than half of the datapoints.%

Overall, \dbg provides more than 30\% speed-up over the baseline on half of the datapoints.
\dbg outperforms or matches skew-aware techniques on nearly all datapoints.
Over the best performing skew-aware technique, \dbg provides the highest performance improvements on the SSSP application, with maximum speed-up of 18.0\% on the \tw dataset.

Structured datasets exhibit high spatio-temporal locality in their original ordering. Thus, any technique that does not preserve the graph structure is likely to yield only a marginal speed-up, if any. 
Among skew-aware techniques, \dbg provides the highest average speed-up of 6.5\% in comparison to -3.7\% for Sort, -2.8\% for HubSort and 5.3\% for HubCluster.

On structured datasets, performance gains from reduction in footprint of hot vertices are negated by disruption in graph structure.
Thus, Sort and HubSort, which preserve graph structure the least, cause slowdown (up to 38.4\%) on more than half of the datapoints. \dbg, in contrast, successfully avoids slowdown on almost all datapoints and causes a marginal slowdown (up to 4.9\%) only on 4 datapoints.

\subsubsection{\textbf{\textit{\dbg vs \gorder}}}
\gorder comprehensively analyzes vertex connectivity to improve cache locality whereas \dbg reorders vertices solely based on their degrees. Thus, it is expected for \gorder to outperform \dbg (and other skew-aware techniques). 
On average, \gorder yields a speed-up of 31.5\% (vs 28.1\% for \dbg) for unstructured datasets and 6.9\% (vs 6.5\% for \dbg) for structured datasets. 

Specifically, difference in speed-ups for \dbg and \gorder is very small for datasets \kr, \tw, \wl and \mpi. These datasets have relatively small clustering coefficient compared to other datasets~\cite{hats}, which makes it difficult for \gorder to approximate suitable vertex ordering. On other datasets, \gorder provides significantly higher speed-ups than any skew-aware techniques. Problematically, \gorder incurs staggering reordering overhead and thus, causes severe slowdowns when accounted for its reordering time (see Sec.~\ref{sec:sw-including-cost}), making it impractical.

\subsubsection{\textbf{\textit{Reordering on No-Skew graphs}}}
In this section, we evaluate the effect of reordering techniques on graph datasets that have no skew. Skew-aware techniques are not expected to provide significant speed-up for these datasets due to lack of skew in their degree distribution. More importantly, these techniques are also not expected to cause any significant slowdown due to a nearly complete lack of locality in the baseline ordering to begin with.

\begin{figure}
    \centering
    \includegraphics{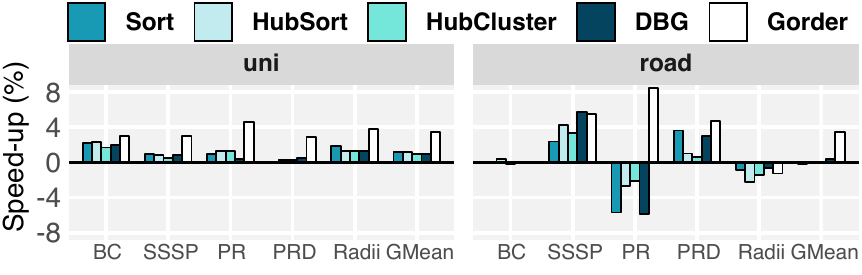}
    \caption{Effect of reordering techniques on graph datasets having no skew.}
    \label{fig:uni}
\end{figure}

Fig.~\ref{fig:uni} shows speed-ups for reordering techniques on two datasets -- {\em uni} and {\em road} -- listed in Table~\ref{tab:no-skew-datasets}. As expected, all skew-aware techniques have a relatively neutral effect, with an average change in execution time within 1.2\% on the {\em uni} dataset and within {0.4\%} on the {\em road} dataset. Meanwhile, Gorder yields slightly more speed-up (3.5\% on both {\em uni} and {\em road} datasets), as it can exploit fine-grain spatio-temporal locality, which is not entirely skew dependent.

\subsection{MPKI Across Cache Levels\label{sec:pull-dominated}}
\begin{figure*}[!t]
    \centering
    \subfloat{\label{fig:mpki-pr-l1}{\transparent{1}\includegraphics[width=1px,height=1px]{pdfs/transparent.png}}}
    \subfloat{\label{fig:mpki-pr-l2}\includegraphics[width=0.99\linewidth]{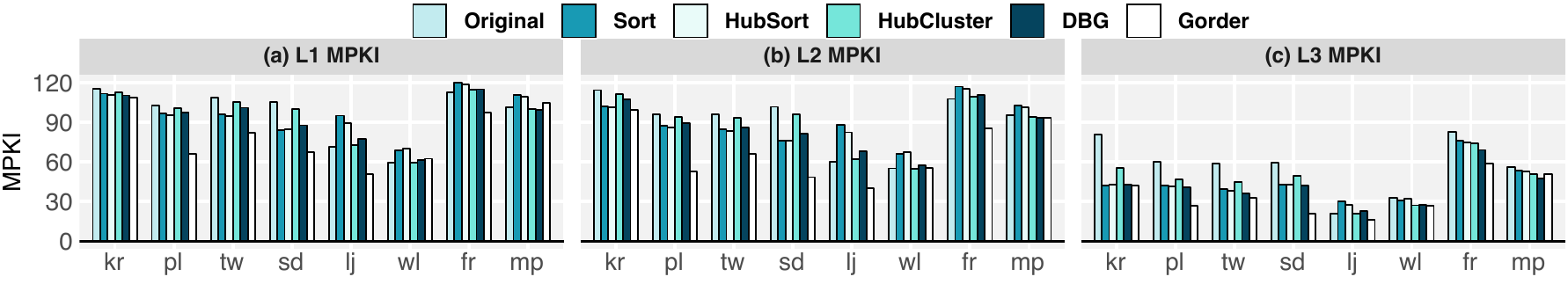}}
    \subfloat{\label{fig:mpki-pr-l3}{\transparent{1}\includegraphics[width=1px,height=1px]{pdfs/transparent.png}}}
    \caption{\label{fig:mpki-pr} Misses Per Kilo Instructions (MPKI) for the PR application across datasets. Lower is better.}
\end{figure*}

In this section, we explain the sources of performance variations for different reordering techniques by analyzing their effects on all three levels of the cache hierarchy.
Fig.~\ref{fig:mpki-pr} plots {\em Misses Per Kilo Instructions (MPKI)} for L1, L2 and L3 cache, measured using hardware performance counters, for the PR application as a representative example.

In the baseline with the original ordering, on all datasets except \lj and \wl, L1 MPKI is more than 100 (\ie at least 1 L1 miss for every 10 instructions on average), which confirms the memory intensive nature of graph applications.
For the original ordering, L2 MPKI is only marginally lower than L1 MPKI across datasets, which shows that almost all memory accesses that miss in the L1 cache also miss in the L2 cache. As L3 cache is significantly larger than L2 cache, L3 MPKI is much lower than L2 MPKI;
nonetheless, L3 MPKI is very high for the original ordering, ranging from 56.2 to 82.9 across large datasets (excluding \lj and \wl).

While all skew-aware techniques target L3 cache, we observe that analyzing the effect of reordering on all three cache levels is necessary to understand application performance.
For example, for \wl{} dataset, Sort yields 5.5\% reduction in L3 MPKI over the baseline and yet causes a slowdown of 5.1\%. 
In fact, the slowdown is caused by 15.3\% and 19.6\% increase in L1 and L2 MPKI, respectively, over the baseline.

\begin{figure*}[!t]
    \centering
    \subfloat{\label{fig:push-false-sharing-orig}{\transparent{1}\includegraphics[width=1px,height=1px]{pdfs/transparent.png}}}
    \subfloat{\label{fig:push-false-sharing-spot}\includegraphics[width=0.995\linewidth]{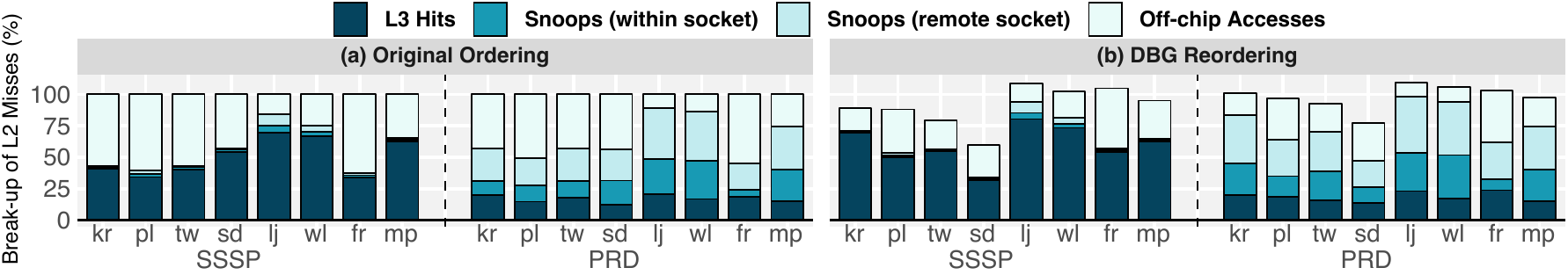}}
    \caption{\label{fig:push-false-sharing} Break-up of L2 misses for the push-dominated applications (SSSP and PRD) for datasets with original and \dbg ordering, normalized to the L2 misses of the original ordering.}
\end{figure*}

All skew-aware techniques are generally effective in reducing L3 MPKI on all datasets but \lj. On unstructured datasets (the left-most four datasets), all skew-aware techniques reduce L1 and L2 MPKI, with the highest reduction on the \sd dataset. Meanwhile, on structured datasets (the right-most four datasets), Sort and HubSort, which do not preserve graph structure, significantly {\em increase} L1 and L2 MPKI (increase of 5.7 to 27.6 over original ordering). In contrast, HubCluster and
\dbg, which largely preserve existing structure, only marginally increase L1 and L2 MPKI (difference of -2.0 to 7.5) on structured datasets.  

Results also show that small datasets are not good candidates for skew-aware reordering. For example, on \lj{} and \wl{} datasets, L3 MPKI is relatively small in the original ordering. As these datasets have relatively fewer number of vertices (5M and 18M vertices, respectively), hot vertices are largely cached in L3, even for the original ordering. Consequently, the opportunity to further exploit skew is small.%

\subsection{Performance Analysis of Push-dominated Applications\label{sec:push-dominated}}

As seen in Fig.~\ref{fig:all-app-perf}, all reordering techniques slowdown the PRD application on many datasets, the cause of which can be attributed to the push-based computation model employed by PRD. 
In push-based computations, when a vertex pushes an update through the out-edges, it generates scattered or irregular write accesses (as opposed to irregular read accesses in pull-based computations). As different threads may concurrently update the same vertex (true sharing) or update different vertices in the same cache block (false sharing), the push-based model leads to read-write or write-write sharing, hence generating on-chip coherence traffic.

Fig.~\ref{fig:push-false-sharing} quantifies coherence traffic on both push-dominated applications -- SSSP and PRD.
The figure shows the break-up of L2 misses into four categories -- L3 Hits (served by L3 without requiring any snoops to other cores), Snoop to other cores within the same socket, Snoops to another socket and off-chip accesses. 
For the first three categories, data is served by on-chip caches whereas for the last category, data is served from the main memory.

The two push-dominated applications have strikingly different fraction of coherence traffic while processing the datasets with the original ordering (middle two stacked bars in Fig.~\stitchref{fig:push-false-sharing}{fig:push-false-sharing-orig}). For SSSP, a relatively small fraction of L2 misses (14.5\% for \lj{} and below 9\% for other datasets) required snoops whereas for PRD, a considerable fraction of L2 misses (from 26.9\% for \fr{} to 69.4\% for \wl{}) required snoops.
 
While processing a vertex using push-based computations, an application pushes updates (writes) to some or all destination vertices of the out-edges. In the case of PRD, it unconditionally pushes an update (\ie a PageRank score) to all destination vertices while processing a vertex. In contrast, SSSP pushes an update to an out-edge only if it finds a shorter path through that edge. Thus, SSSP has much fewer number of irregular writes, and in turn, less coherence traffic, in comparison to PRD.

\begin{figure*}[!t]
    \centering
    \includegraphics[width=1\linewidth]{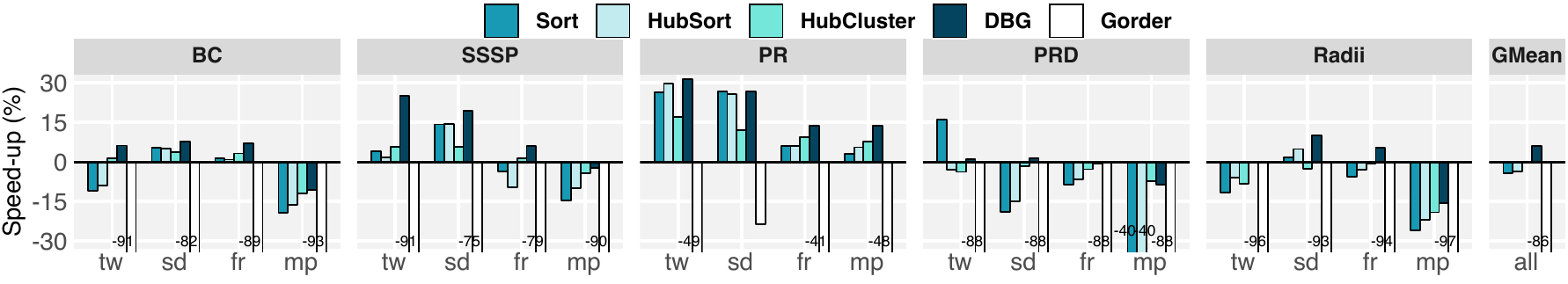}
    \caption{\label{fig:all-cost-perf}Net speed-up for software reordering techniques over the baseline with original ordering of datasets. GMean shows geometric mean across speed-ups for all five applications on four datasets.}
\end{figure*}

\begin{figure*}[!t]
    \centering
    \includegraphics[width=1\linewidth]{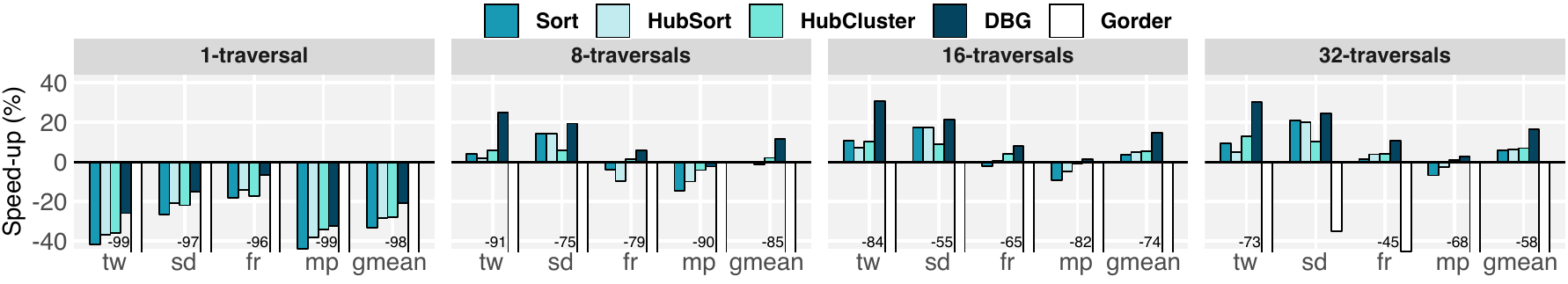}
    \caption{\label{fig:net-cost-sssp}Net speed-up for reordering techniques over the baseline with no reordering for SSSP with different number of traversals.}
\end{figure*}

Fig.~\stitchref{fig:push-false-sharing}{fig:push-false-sharing-spot} shows a similar break-up for SSSP and PRD on the datasets after \dbg reordering.
For PRD, \dbg consistently reduces off-chip accesses (top stacked bar) across datasets, thus, a significantly higher fraction of requests are served by on-chip caches. However, most of these requests (37.8\% to 77.0\% of L2 misses) incur a snoop latency. For example, for \dbg, while processing the \pl{} dataset, 65.4\% (vs 49.2\% for the original ordering) of L2 misses are served by on-chip caches (bottom three stacked bars combined). However, most of these on-chip hits required snooping to other cores, incurring high access
latency. Specifically, only 18.9\% (vs 14.8\% for the original ordering) of total L2 misses are served without requiring snooping. For most of the datasets, increase in L3 hits (\ie no snooping) due to \dbg is relatively small despite a significant reduction in off-chip accesses, which explains the marginal speed-up for \dbg for the PRD application (Fig.~\ref{fig:all-app-perf}).

For SSSP, most of the savings in off-chip accesses directly translate to L3 hits (\ie no snooping) as the application does not exhibit high amount of coherence traffic even in the baseline. Thus, \dbg is highly effective on SSSP, despite being dominated by push-based computations.

\subsection{Performance Including Reordering Time\label{sec:sw-including-cost}}

\setlength\tabcolsep{4pt}
\begin{table}[!t]
    \centering
    \begin{tabularx}{1\linewidth}
        {|>{\centering\arraybackslash\hsize=0.15\hsize}X| 
          >{\raggedleft\arraybackslash\hsize=0.16\hsize}X|
          >{\raggedleft\arraybackslash\hsize=0.16\hsize}X|
          >{\raggedleft\arraybackslash\hsize=0.21\hsize}X|
          >{\raggedleft\arraybackslash\hsize=0.16\hsize\columncolor{lightgray}}X|
          >{\raggedleft\arraybackslash\hsize=0.16\hsize}X|
        }
        \hline
            Dataset & \centering Sort    & \centering HubSort   & \centering HubCluster    & \centering \dbg    & \centering \gorder \tabularnewline \hline \hline
            \tw  &  3.3  & 2.4  & 3.5  & 1.9  & 258.6 \tabularnewline \hline
            \sd  &  3.7  & 3.0  & 5.0  & 2.4  & 112.2 \tabularnewline \hline
            \fr  &  8.6  & 7.4  & 4.7  & 3.2  & 254.9 \tabularnewline \hline
            \mpi  &  18.2  &    10.3  &    7.5  & 4.4  & 1359.4 \tabularnewline \hline

    \end{tabularx}
    \caption{Minimum number of iterations of PR application needed to amortize the reordering time of different reordering techniques.}
    \label{tab:min-iter-amortize}
\end{table}

Fig.~\ref{fig:all-cost-perf} shows end-to-end application speed-up for different reordering techniques after accounting for the reordering time. Due to space constraints, we show only four datasets (two largest unstructured and two largest structured datasets). 

\gorder{}, while more effective at improving application speed-up (Fig.~\ref{fig:all-app-perf}), when accounted for its reordering time, causes severe slowdowns (up to 96.5\%) across datasets, corroborating prior work~\cite{hubcluster}. In contrast, all skew-aware techniques provide a net speed-up on at least some of the datapoints.

\dbg outperforms all prior techniques on 17 out of 20 datapoints. \dbg provides a net speed-up (up to 31.4\%) on 14 out of 20 datapoints, even after accounting for its reordering time. On the remaining 6 datapoints, \dbg reduces slowdown when compared to prior techniques, with maximum slowdown of 15.6\% for the Radii application on the \mpi dataset and below 10\% for others.
In contrast, existing skew-aware techniques cause slowdown of up to 40.2\% on half of the datapoints. Overall, DBG is the only technique that yields an average net speed-up (6.2\%) by providing high performance while incurring low reordering overhead.

We next study how 
long it takes to amortize the reordering cost for an iterative application (PR) and a root-dependent traversal application (SSSP).

\subsubsection{\textbf{\textit{Amortization point for PR}}}
The PR application has the largest runtime among all five applications for any given dataset, thus all skew-aware techniques are highly effective for the PR application and yield a net speed-up on all four datasets. 
Averaging across four datasets for the PR application, \dbg outperforms all reordering techniques with 21.2\% speed-up vs 15.1\% for Sort, 16.3\% for HubSort, 11.6\% for HubCluster and -41.3\% for~\gorder. %

Table~\ref{tab:min-iter-amortize} lists the minimum number of iterations needed for the PR application to amortize the cost of different reordering techniques. For all four datasets, \dbg is quickest in amortizing its reordering time, providing a net speed-up for all four datasets after just 2-5 iterations. 

\subsubsection{\textbf{\textit{Amortization point for SSSP}}}
We now evaluate net performance sensitivity to the number of successive graph traversals for different techniques for the SSSP application.
The runtime for root-dependent applications depends on the number of traversals (or queries) performed from different roots. The exact number of traversals required depends on the specific use case. Thus, we perform a sensitivity analysis by varying the number of traversals from 1 to 32 in multiples of 8. 

As shown in Fig.~\ref{fig:net-cost-sssp}, with the increase in the number of traversals, performance for each technique also increases, as the reordering needs to be applied only once and its cost is amortized over multiple graph traversals. Thus, a single traversal is the worst-case scenario, with all techniques causing slowdown due to their inability to amortize the reordering cost. Of all the techniques, DBG causes the minimum slowdown (20.6\% on average vs 27.7\% for the next best) and is the quickest in amortizing the reordering cost, providing an average speed-up of 11.5\% (vs 2.1\% for the next best) with as few as 8 graph traversals.

\section{Related Work}
\label{related}

A significant amount of research has focused on designing high performance software frameworks for graph applications (e.g., ~\cite{ligra, galois,gap, graphmat, graphlab, graphchi}). In this section, we highlight the most relevant works that focus on improving cache efficiency for graph applications. %

\noindentsectiontitle{\bf Graph Slicing:}
Researchers have proposed graph slicing that slices the graph in LLC-size partitions and processes one partition at a time to reduce irregular memory accesses~\cite{fc,graphicionado,graphmat}.
While generally effective, slicing has two important limitations. First, it requires invasive framework changes to form the slices (which may include replicating vertices to avoid out-of-slice accesses) and manage them at runtime. Secondly, for a given cache size, the number of slices increases with the size of the graph, resulting in greater processing overheads in creating and maintaining partitions for larger graphs.
In comparison, \dbg only requires a preprocessing pass over the graph dataset to relabel vertex IDs and does not require any change in the graph algorithms.

\noindentsectiontitle{\bf Traversal Scheduling:}
A recent work proposed Bounded Depth-First Scheduling (BDFS) to exploit cache locality for graphs exhibiting community structure~\cite{hats}.
However, the software implementation of BDFS introduces significant book-keeping overheads, causing slowdowns despite improving cache efficiency. To avoid software overheads, the authors propose an accelerator that implements BDFS scheduling in hardware. In comparison, \dbg is a software technique that can improve application performance without any additional hardware support.

\noindentsectiontitle{\bf Hardware Cache Management:} Many researchers have proposed domain-agnostic hardware cache management schemes to reduce cache thrashing~\cite{ship,sampler,hawkeye,perceptron, leeway}. These hardware schemes improve cache utilization by keeping high reuse cache blocks in the LLC based on inter-block reuse and thus are orthogonal to DBG (and other skew-aware software techniques) that improve cache block utilization by focusing on intra-block reuse. 

A recent work have proposed a domain-specialized cache management scheme for graph analytics~\cite{GRASPFalduPACT19}, which requires hot vertices to be located in a contiguous memory region. DBG is inherently compatible with such a scheme. DBG can also facilitate a wider adoption of such a hardware scheme over more complex reordering techniques such as Gorder (assuming the cost of complex techniques can be tolerated). To achieve this, DBG can be applied after applying Gorder on an original dataset to further reorder vertices, which results in a vertex order that retains most of the Gorder ordering while also segregating hot vertices in a contiguous region. The combined reordering retains most of the performance of Gorder due to its structure preserving property 
(e.g., Gorder+DBG achieves 17.2\% average application speed-up across 40 datapoints vs 18.6\% for Gorder alone), while making dataset compatible with such a hardware scheme to gain further performance improvement.

\section{Future Work}
\label{future}
We believe that there are two potential future research directions for the wider domain of vertex reordering in the context of graph generation and dynamic graphs.

\subsection{Integrating Reordering Techniques with Graph Generation}
In this work, we assumed that the graph datasets are readily available and thus assumed that spatio-temporal locality in real-world datasets (specifically for the structured datasets) exist without any overhead. In practice, such ordering may be a positive side effect of dataset generation algorithm (e.g., crawling webpages in certain order) or it may have been achieved by post-processing a dataset (e.g., graph datasets available from The Laboratory for Web Algorithmics have been ordered with the Layered Label Propagation technique~\cite{llp}). Thus, there exist an opportunity to integrate skew-aware reordering techniques with the dataset generation process in order to avoid regenerating CSR-like structure post reordering, which dominates the reordering cost. At the very least, cost of reordering should be compared to the cost of a post-processing technique used over the raw dataset to understand the cost-benefit trade-offs of techniques from different domains.

\subsection{Amortizing Reordering Costs on Dynamic Graphs}
In this work, we assumed that graphs are static and thus, have evaluated a net speed-up conservatively assuming only one graph application (or query) over the reordered dataset (Fig.~\ref{fig:all-cost-perf}). In practice, a graph may evolve over time and a stream of graph updates (i.e., addition or removal of vertices or edges) are interleaved with graph-analytic queries. For such a deployment, graph reordering may provide an even greater benefit as the reordering cost can be amortized not only over multiple graph traversals of a single query, but also over multiple graph queries. Intuitively, addition or removal of some vertices or edges in a large graph would not lead to a drastic change in the degree distribution, and thus unlikely to change which vertices are classified hot in a short time window. Therefore, reordering techniques may need to be re-applied at large periodic intervals (\ie after a series of updates has been made to a graph) to improve cache behavior,
amortizing the cost of reordering over multiple graph queries performed in a given interval.

\section{Conclusion}
\label{conclusion}

In this work, we study skew-aware reordering techniques that seek to improve cache efficiency for graph analytics. Our work demonstrates the inherent tension between reducing the cache footprint of hot vertices and preserving original graph structure, which limits the effectiveness of existing skew-aware reordering techniques. To overcome the limitations of existing techniques, we propose Degree-Based Grouping (DBG), a novel lightweight reordering technique that employs
coarse-grain reordering to preserve graph structure while reducing the cache footprint of hot vertices. DBG outperforms all existing skew-aware techniques, yielding an average speed-up of 16.8\% vs 11.6\% for the best-performing skew-aware technique.

\section*{Acknowledgment}
We thank Artemiy Margaritov, Amna Shahab and the anonymous reviewers for their valuable feedback on an earlier draft of this work. This work was supported in part by a research grant from Oracle Labs.

\bibliographystyle{IEEEtran}
\bibliography{99-ref}

\end{document}